\documentclass[aps,prc,superscriptaddress,nofootinbib,showpacs,floatfix,singlecolumn]{revtex4}
\usepackage{url}
\usepackage{cancel}
\usepackage[colorlinks,linkcolor=blue,citecolor=blue,filecolor=black,urlcolor=blue]{hyperref}
\usepackage{epsfig,graphics}
\usepackage{graphicx}
\usepackage{dcolumn}
\usepackage{bm}
\usepackage[usenames]{color}
\usepackage{amssymb}
\usepackage{amsmath}
\usepackage{multirow}
\usepackage{float}
\usepackage{harpoon}
\usepackage{MnSymbol}
\usepackage{appendix}
\usepackage{color}
\usepackage{hyperref}
\usepackage{cleveref}

\newcommand{\sqrtsnn}{\mbox{$\sqrt{s_{\mathrm{NN}}}$}}
\newcommand{\pT} {p_{\mathrm{T}}}
\newcommand{\lr}[1]{\left\langle #1\right\rangle}

\newcommand{\nall}{N_{\mathrm{hadron}}}
\newcommand{\npart}{N_{\mathrm{part}}}
\newcommand{\rhon}[1] {\rho(v_{ #1 }^2,[\pT])}
\newcommand{\rhone}[1] {\rho(\varepsilon_{ #1 }^2,S/A)}
\newcommand{\cov}[1] {\lr{v_{ #1 }^2\delta\pT}}
\newcommand{\cove}[1] {\lr{\varepsilon_{ #1 }^2\delta \frac{S}{A}}}
\newcommand{\var}[1] {\mathrm{var}(v_{ #1 }^2)}
\newcommand{\vare}[1] {\mathrm{var}(\varepsilon_{ #1 }^2)}
\newcommand{\varp} {\mathrm{var}([\pT])}
\newcommand{\vares} {\mathrm{var}(S/A)}

\newcommand{\ptk}[1] {p_{\mathrm{T},#1}}

\usepackage{CJK}
\hfuzz=\maxdimen
\begin{document}
\title{Probing nuclear quadrupole deformation from correlation of elliptic flow and transverse momentum in heavy ion collisions}
\newcommand{\sbu}{Department of Chemistry, Stony Brook University, Stony Brook, NY 11794, USA}
\newcommand{\bnl}{Physics Department, Brookhaven National Laboratory, Upton, NY 11976, USA}
\author{Jiangyong Jia}\email[Correspond to\ ]{jiangyong.jia@stonybrook.edu}\affiliation{\sbu}\affiliation{\bnl}
\author{Shengli Huang}\affiliation{\sbu}
\author{Chunjian Zhang}\affiliation{\sbu}
\begin{abstract}
In heavy ion collisions, elliptic flow $v_2$ and radial flow, characterized by event-wise average transverse momentum $[p_{\mathrm{T}}]$, are related to the shape and size of the overlap region, which are sensitive to the shape of colliding atomic nuclei. The Pearson correlation coefficient between $v_2$ and $[p_{\mathrm{T}}]$, $\rho_2$, was found to be particularly sensitive to the quadrupole deformation parameter $\beta$ that is traditionally measured in low energy experiments. Built on earlier insight that the prolate deformation $\beta>0$ reduces the $\rho_2$ in ultra-central collisions (UCC), we show that the prolate deformation $\beta<0$ enhances the value of $\rho_2$. As $\beta>0$ and $\beta<0$ are the two extremes of triaxiality, the strength and sign of $v_2^2-[p_{\mathrm{T}}]$ correlation can be used to provide valuable information on the triaxiality of the nucleus. Our study provide further arguments for using the hydrodynamic flow as a precision tool to directly image the deformation of the atomic nuclei at extremely short time scale ($<10^{-24}$s).
\end{abstract}

\pacs{25.75.Gz, 25.75.Ld, 25.75.-1}
\maketitle

\section{Introduction}\label{sec:1}
Heavy-ion collisions at RHIC and the LHC produce a Quark-Gluon Plasma (QGP) whose space-time evolution is well described by relativistic viscous hydrodynamics~\cite{Gale:2013da,Heinz:2013th,Florkowski:2017olj,Busza:2018rrf}. Driven by large pressure gradients, the QGP undergoes collective, Hubble-like expansion in the transverse plane, converting spatial non-uniformities in the initial state into the collective radial and azimuthally anisotropic flow in the final state. We quantify such collectivity via a Fourier expansion of particle distribution in azimuth $\phi$ and transverse momentum $\pT$: $\frac{d^2N}{\pT d\pT d\phi} = N(\pT) \left[1+2\sum_{i=1}^{\infty} v_n(\pT) \cos n(\phi-\Psi_n)\right]$, where $v_n$ and $\Psi_n$ represent the amplitude and phase of the $n^{\mathrm{th}}$-order anisotropic flow, and the slope of the particle spectrum $N(\pT)$ characterizes the magnitude of the radial flow. The strength of radial and anisotropic flow depends on the initial state: a compact source generates a stronger radial flow reflected by a flatter spectrum, and a more eccentric shape of the source leads to larger anisotropic flow. They are also sensitive to the transport properties of the QGP such as shear and bulk viscosity. Recent state-of-the-art comparisons between hydrodynamics and precision $v_n$ and $\pT$ spectra data provided quantitative constraint on both the properties of the medium, as well as the density fluctuations in the initial state~\cite{Bernhard:2016tnd,Everett:2020xug,Nijs:2020ors}.

It is a well established fact that the $v_n$ are driven by hydrodynamic response to the initial eccentricity $\varepsilon_n$ of QGP medium, which can be estimated from the position $(r,\phi)$ of participating nucleons, $\varepsilon_n = |\int r^n e^{-in\phi} drd\phi/\int r^ndr|$~\cite{Teaney:2010vd}.  Model calculations show that the $v_n$ are approximately proportional to $\varepsilon_n$ for $n=2$ and 3~\cite{Niemi:2012aj}. The radial flow, characterized by the average transverse momentum in each event $[\pT]$,  reflects the hydrodynamic response to the fluctuation in the overall size of the overlap region $R$. In particular, events with similar total energy, but smaller transverse size in the initial state is expected to have stronger radial expansion and larger $[\pT]$~\cite{Bozek:2012fw,Chatterjee:2017mhc}. Therefore, the event-by-event fluctuation in the shape and size of the QGP can be inferred from the fluctuations of the $v_n$ and $[\pT]$ in the final state. 

In collisions of spherical nuclei, the shape and size of the QGP are controlled by the impact parameter. For deformed nuclei, however, they depend also on the quadrupole deformation parameters $\beta$ and $\gamma$ as part of the Woods-Saxon density function:
\begin{align}\label{eq:1}
\rho(r,\theta,\phi)=\frac{\rho_0}{1+e^{(r-R(\theta,\phi)/a)}},\;\; R(\theta,\phi) = R_0\left(1+\beta [\cos \gamma Y_{2,0}+ \sin\gamma Y_{2,2}]\right).
\end{align}
where $\rho_0$ is the density at the center of the nucleus, and nuclear radius $R_0$ are nuclear radius and $a$ is the skin depth. The quadrupole-shaped nuclear surface $R(\theta,\phi)$ is expanded in terms of spherical harmonics in real form. The three components $Y_{2,-1},Y_{2,1}$ and $Y_{2,-2}$ are customarily used to defined the body-fixed x-y-z frame, leaving $Y_{2,0}$ and $Y_{2,2}$ as the only relevant degrees of freedom. The mixing angle $0\leq\gamma\leq\pi/3$ controls the triaxiality or the three radii $R_a,R_b,R_c$ of the nucleus in the body-fixed frame, with $\gamma=0$, $\gamma=\pi/3$, and $\gamma=\pi/6$ corresponding to prolate ($R_a=R_b<R_c)$, oblate ($R_a<R_b=R_c$) or maximum triaxiality ($R_a<R_b<R_c$ forming an arithmetic sequence). Note that the oblate shape can be specified either as $\beta,\gamma=\pi/3$ or equivalently as $-\beta$ and $\gamma=0$. In this present study we only consider prolate and oblate configuration for which we can keep $\gamma=0$ and let $\beta$ change sign. Nevertheless, the comparison of results between prolate and oblate deformation still provides critical information on the influence of triaxiality. 

Most stable nuclei in their ground states are quadrupole-deformed and has a non-zero $\beta$. The values of $\beta$ are obtained from measurement of rotational spectra of nuclear excited state or the electric quadrupole moments from hyperfine splitting of atomic spectral line~\cite{Moller:2015fba}. Due to the random orientation of the colliding nuclei, quadrupole deformation enhances the event-by-event fluctuations of the $\varepsilon_2$ and $v_2$. This point was investigated extensively and could explain the ordering of the $v_2$ data in ultra central collisions (UCC) of different collision systems~\cite{Acharya:2018ihu,Sirunyan:2019wqp,Aad:2019xmh}. Model studies show that the mean square fluctuation of $\varepsilon_2$ and $v_n$ depends quadratically on $\beta$, $\varepsilon_2\{2\}^2\equiv\lr{\varepsilon_2^2}=a'+b'\beta^2$ and $v_2\{2\}^2\equiv\lr{v_2^2}=a+b\beta^2$~\cite{Giacalone:2021udy,Jia:2021tzt}. Interestingly, the response coefficients for the $\beta$-independent and $\beta$-dependent components of $v_2$ and $\varepsilon_2$ are not the same, i.e. $a/a'\neq b/b'$~\cite{Giacalone:2021udy}. This opens up the possibility to test hydrodynamics using $\beta$ as a new control variable, i.e. by comparing nucleus with similar mass number but different $\beta$. Recently, quadrupole deformation was also predicted to have strong influence on correlated fluctuation between $v_2$ and $[\pT]$~\cite{Giacalone:2019pca,Giacalone:2020awm}, quantified by a three-particle correlator~\cite{Bozek:2016yoj}:
\begin{align} \label{eq:2}
\rhon{n}  =\frac{\lr{v_n^2\delta\pT}}{\sqrt{\lr{\left(\delta v_n^2\right)^2}\lr{\delta \pT\delta \pT}}}
\end{align}
where $\delta \pT = \pT-[\pT]$ and the ``$\lr{}$'' denotes averaging over all pairs or triplets for events with similar particle multiplicity. This observable can be approximated by an analogous quantity calculated from the initial state~\cite{Schenke:2020uqq}:
\begin{align} \label{eq:3}
\rhone{n}=\frac{\left\langle{\delta} \varepsilon_{n}^{2} {\delta}(\frac{S}{A})\right\rangle}{\sqrt{\left\langle\left({\delta} \varepsilon_{n}^{2}\right)^{2}\right\rangle\left\langle({\delta}(\frac{S}{A}))^{2}\right\rangle}}
\end{align}
where the $S/A$ is the initial entropy density in the transverse plane. The $\rhon{2}$ is positive for spherical systems, but for nuclei with large prolate deformation, the $\rhon{2}$ values in UCC is predicted to be negative. This is because selection of central events in U+U collisions enhances body-body events, which have large $\varepsilon_2$ and $R$, therefore large $v_2$ and smaller $[\pT]$~\cite{Giacalone:2019pca}. Preliminary results from the STAR Collaboration support this interpretation~\cite{jjia}. Therefore using the well-tuned hydrodynamics model as a precision tool and together with the experimental measurements of $v_2$ and $v_2-[\pT]$, we could provide quantitative constraint on the shape of the nuclei at a time scale of $10^{-24}$s, which is much shorter than that involved in the low energy nuclear structure measurements. 

In this paper, we study the influence of quadrupole deformation to correlations between $v_2$ and $[\pT]$. In particular, we clarify the relation between the initial-state estimator Eq.~\ref{eq:3} and final-state experimental observable Eq.~\ref{eq:2}. We perform this study using the ``a multi-phase transport model'' (AMPT), which is a realistic yet computationally efficient way to implement hydrodynamic response~\cite{Lin:2004en}. We carry out simulation of Au+Au and U+U collisions at top RHIC energy with different $\beta$ values, ranging from prolate to oblate configurations. We found that large oblate deformation gives rise to an enhanced positive $\rhon{2}$ value in UCC region. This is because the body-body collisions of oblate nuclei have large $\varepsilon_2$ and but smaller $R$, therefore large $v_2$ and $[\pT]$, exactly opposite to the influence of prolate deformation. We also quantified the effects of volume fluctuations based on realistic centrality definition matching to experimental acceptance, and they are found to be minimized in the UCC collisions. Future detailed model-data comparison will firm up the relation between initial and final state and provide some useful constrains on the shape of deformed nuclei. Just before the submission of this work, a study of $\rhon{2}$ in Au+Au based on AMPT model appeared, whose focus however was not on the influence of deformation~\cite{Magdy:2021ocp}. 

\section{analysis}\label{sec:2}
We calculate Pearson correlation coefficient $\rhon{n}$ within the usual multi-particle cumulant framework employed by the experimental data analysis as detailed in \cite{Zhang:2021phk}, which we just briefly summarize here. The numerator of $\rhon{n}$ is obtained by averaging over unique triplets in each event, and then over all events in an event class~\cite{Bozek:2016yoj,Bozek:2020drh}:
\begin{align}\label{eq:4}
\lr{v_n^2\delta\pT} &= \lr{\frac{\sum_{i,j,k, i\neq j\neq k} e^{in(\phi_i-\phi_j)}(\ptk{k}-\lr{ [\pT]})} {\sum_{i,j,k, i\neq j\neq k}}}
\end{align}
where the indices $i$, $j$ and $k$ loop over distinct particles to account for all unique triplets, and the $\lr{}$ denotes average over events. In this analysis, we use all particles within $|\eta|<2$ and $0.2<\pT<2$ GeV for best statistical precision. However, we also checked the influence of short-range ``non-flow'' correlations in the context of the so-called two-subevent and three-subevent methods~\cite{Jia:2017hbm} by introducing pseudorapidity gaps between the particles in each triplet. We conclude that non-flow effects are negligible (see Appendix~\ref{sec:app1}).

The $[\pT]$ variance in the denominator of ${\rho}_n$ is obtained using usual two-particle $\pT$ correlations~\cite{Abelev:2014ckr}, 
\begin{align}\label{eq:5}
\varp\equiv \lr{\delta \pT\delta \pT}= \lr{\frac{\sum_{i,j, i\neq j}(\ptk{i}-\lr{[\pT]})(\ptk{j}-\lr{[\pT]})}{\sum_{i,j,i\neq j}}}\;.
\end{align}
The flow variance is calculated in terms of two-particle cumulants $c_n\{2\}$ and four-particle cumulants $c_n\{4\}$ following Ref.~\cite{ATLAS:2021kty}: $\var{n} \equiv \lr{\left(\delta v_n^2\right)^2}= \lr{v_n^4}-\lr{v_n^2}^2 =  c_n\{4\} + c_n\{2\}^2$.

We calculate the initial-state estimator in Eq.~\ref{eq:3} from the transverse distribution of participating nucleons. Here we substitute the total entropy $S$ with $\npart$ with the assumption that $S\propto\npart$. Following the recommendation of Ref.~\cite{Schenke:2020uqq}, we define the overlap area $A$ as 
\begin{align}\label{eq:7}
A=2\pi\sqrt{\sigma_x^2\sigma_y^2} 
\end{align}
where the $\sigma_x$ and $\sigma_y$ are the RMS width of participant distribution along the short and long principal axes of the event-by-event ellipse. This definition was shown to give very good correlation with the event-by-event $[\pT]$. 

We perform the calculation of all these observables within the AMPT transport model~\cite{Lin:2004en}. The model starts with Monte Carlo Glauber initial conditions. The system evolution is modeled with strings that first melt into partons, followed by elastic partonic scatterings, parton coalescence, and hadronic scatterings. The collectivity is generated mainly through elastic scatterings of partons, which leads to an emission of partons preferable along the gradient of the initial-state energy density distribution, in a manner that is similar to hydrodynamic flow. Following Refs~\cite{Ma:2014pva,Bzdak:2014dia,Nie:2018xog}, we use the AMPT model v2.26t5 with string-melting mode and partonic cross section of 3.0~mb, which we check reasonably reproduce Au+Au $v_2$ data at RHIC. The Woods-Saxon parameters in the AMPT are chosen to be $R_0=6.81$~fm and $a=0.535$~fm for U+U similar to ~\cite{Heinz:2004ir} and $R_0=6.37$ fm and $a=0.54$ fm for Au+Au~\cite{Raman:1201zz}. For the study on the $\beta$ dependence, we simulate collisions at $\sqrtsnn=200$ GeV for U+U with $\beta$=0,-0.15,0.22,$\pm$0.28,0.34 and 0.4 and for Au+Au with $\beta$=0 and -0.13, which will allow us to obtain the parametric dependence of various flow observables on $\beta$. This list includes the default $\beta$ value of 0.28 for U+U and -0.13 for Au+Au from the most recent table of nuclear deformations~\cite{Moller:2015fba}.

Our main analysis is performed using all hadrons with $0.2<\pT<2$ GeV and $|\eta|<2$, and the event centrality is defined using either $\npart$ or inclusive hadron multiplicity in $|\eta|<1$, $\nall$. The value of $\nall$, which include both charged and neutral particles, is about three times of the charged hadron multiplicity density, i.e. $\nall\approx 3 dN_{\mathrm{ch}}/d\eta$. It is known that the multi-particle correlations are sensitive to centrality/volume fluctuations, i.e. the centrality measured based on the final-state particle multiplicity in a $\eta$ range is subject to smearing due to fluctuations in the particle production process~\cite{Skokov:2012ds,Zhou:2018fxx}. Since the $v_n$ and $[\pT]$ values vary with centrality, the smearing in centrality can lead to additional fluctuations of shape and size of the overlap region. Indeed, significant differences in $\rhon{n}$ were observed in the ATLAS measurement, when the results are compared between centrality defined in mid-rapidity and centrality defined in the forward rapidity~\cite{ATLAS:2021kty}. To quantify the volume fluctuation effects, we perform a separate analysis with alternative selection on particle of interest and particles used to define centrality. Details of this study can be found in the Appendix~\ref{sec:app1}.

\section{results} \label{results}
\begin{figure}[h!]
\begin{center}
\includegraphics[width=1\linewidth]{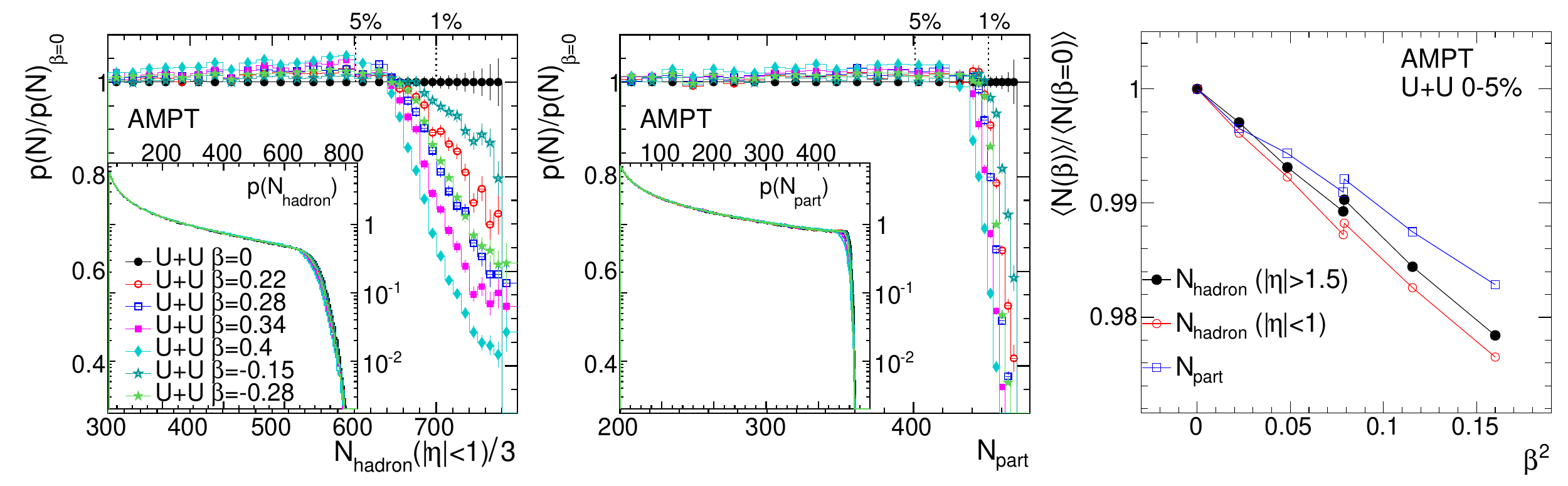}
\end{center}
\caption{\label{fig:0} The distributions and ratios of hadron multiplicity $\nall$ in $|\eta|<1$ scaled by 1/3 (left) and $\npart$ (middle) in U+U collisions with different quadrupole deformation parameter $\beta$. The vertical lines indicate locations of 1\% and 5\% for $\beta=0$ case. The right panel shows the average $\nall$ or $\npart$ in the top 0--5\% of events selected according to $\nall$ or $\npart$ as a function $\beta^2$, they are normalized the corresponding values in $\beta=0$. Also in the right panel, the data point for $\beta=-0.28$ is shifted slightly to the right to distinguish from the data point for $\beta=0.28$.}
\end{figure}

In collisions of spherical nuclei, the multiplicity distributions $p(\nall)$ or $p(\npart)$ are controlled by the impact parameter. In the presence of deformation, these distributions are expected to be smeared and broadened. The insert small panels in Fig.~\ref{fig:0} show the multiplicity distribution in U+U collisions with different $\beta$ values. These distributions are divided by those for $\beta=0$ and the results are shown in the corresponding main panels. We see a clear reduction of the ratio for large $\nall$ or large $\npart$ values, and an increase in other regions, confirming the broadening of multiplicity distribution for non-zero $\beta$. We also observe that the ratios are very similar between $\beta=0.28$ and -0.28. The influence of quadrupole deformation is clearly visible only in the most central 0--5\% region. To quantify the influence of quadrupole deformation on the multiplicity distributions, we calculate $\lr{\nall}$ and $\lr{\npart}$ in the top 0--5\%, normalized by the values for $\beta=0$, and plot the results in the right panel of Fig.~\ref{fig:0}. We observe a linear decrease of the $\lr{\nall}$ or $\lr{\npart}$ as a function of $\beta^2$, implying that the multiplicity smearing is same for prolate and oblate deformation with same $|\beta|$. The decrease in the average multiplicity is only around 1\% for the realistic deformation value of $\beta=0.28$, so it is a modest effect but should be visible for Ru+Ru and Zr+Zr isobar collision systems at RHIC~\cite{Xu:2021vpn}. Parameterizing this dependence by $\lr{\npart}=a_0+b_0\beta^2$, one can see that the coefficients $a_0$ and $b_0$ can be determined from two isobar systems with known $\beta^2$, which can then be used to gauge the $\beta$ value of other systems with same mass number. We also notices that the extent of decrease also depends on the definition of multiplicity variable. In general, the effect is smallest when $\npart$ is used, and is largest when $\nall$ is defined around mid-rapidity. The latter is consistent with the finding by the ATLAS Collaboration that centrality resolution is worse at mid-rapidity than the forward rapidity~\cite{Aaboud:2019sma}.  

We would like to point out that in the presence of large deformation, the total volume of the nucleus increases slightly. For the largest value considered, $\beta=0.4$, the ratio to the original volume is $1+\frac{3}{4\pi}\beta^2+\frac{\sqrt{5}}{28\pi^{3/2}}\cos(3\gamma)\beta^3=1.029+0.0006\cos(3\gamma)$. In order to keep the overall volume fixed, it would require about 1\% decrease of the $R_0$. We have performed a separate Glauber model investigation on the impact of this change of $R_0$, which is found to have negligible influence on the slopes shown in the right panel of Fig.~\ref{fig:0}.

Having established the fact that the impact of $\beta$ on multiplicity and centrality distributions is small, we are ready to discuss the $v_n-[\pT]$ correlations. In order to have a clear connection between initial-state deformation and final-state correlations, we always compare the results calculated using final-state particles via Eq.~\ref{eq:2} with the estimators based on the initial-state Glauber geometry via Eq.~\ref{eq:3}. 

\begin{figure}[h!]
\begin{center}
\includegraphics[width=1\linewidth]{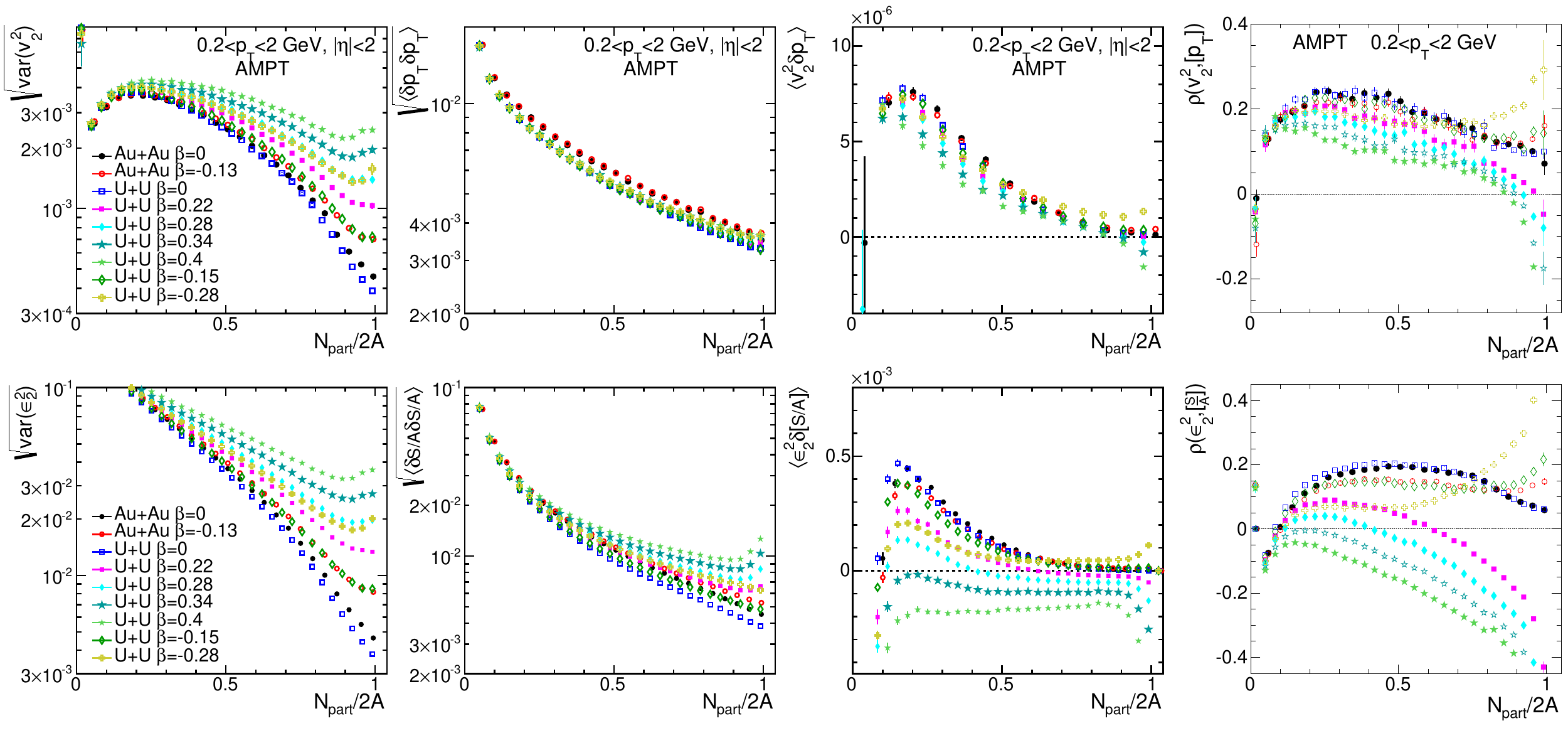}
\end{center}
\caption{\label{fig:1} The $\npart$ dependence of $v_2^2$ variance (top-left), $[\pT]$ variance (2nd from top-left), flow-$[\pT]$ covariance $\cov{2}$ (3rd from top-left) and Pearson coefficients $\rhon{2}$ (top-right) in Au+Au and U+U collisions with different deformation parameter $\beta$. The bottom panels show the corresponding initial-state estimators in the Glauber model. The event class used for averaging is based on $\npart$.}
\end{figure}

The top panels of Fig.~\ref{fig:1} show the components of the Pearson correlation coefficient $\rhon{2}$: the $\sqrt{\var{2}}$, $\sqrt{\varp}$, $\cov{2}$ and $\rhon{2}$ calculated using final-state hadrons in Au+Au and U+U collisions with different $\beta$. The corresponding quantities from the initial state are shown in the bottom panels. The $\sqrt{\var{2}}$ show very strong dependence on $\beta$, as argued in Ref.~\cite{Giacalone:2021udy}, reflecting mainly a linear response to the eccentricity fluctuations $\sqrt{\vare{2}}$ in the initial state. The $\varp$ only show a very modest dependence on quadrupole deformation, it increases by about 10\% in U+U collisions from $\beta=0$ to $\beta=0.4$. In contrast, the initial-state estimator $\vares$ shows a much stronger increase in the presence of deformation. The reason is that the AMPT model fails to describe the radial flow and its fluctuations~\cite{Zhang:2021vvp}. In fact we found AMPT underestimates the $\sqrt{\varp}$ data~\cite{Adam:2019rsf} for Au+Au at RHIC by more than factor of two, and this problem is there for all recent versions of AMPT. This failure implies that radial flow response to the size of the system is too weak in AMPT and might preclude a quantitative comparison with the experimental measurement of $\rhon{2}$. We also note that the both $\sqrt{\vare{2}}$ and $\sqrt{\vares}$ are similar between $\beta=-0.28$ and $\beta=0.28$, implying they are mostly even functions of $\beta$~\footnote{However in the UCC region (the last couple of points), we see that the values of $\sqrt{\vare{2}}$ are slightly larger while the values of $\sqrt{\vares}$ are smaller for $\beta=-0.28$ than for $\beta=0.28$. This reverse ordering is a bit more clear when events are binned based on $\nall$ (see Fig.~\ref{fig:app3})}. Both $\cov{2}$ and its initial-state counterpart $\cove{2}$ show strong yet non-trivial dependence on $\beta$. For prolate deformation $\beta>0$, the covariance decreases with increasing $\beta$ values. However, for oblate deformation $\beta<0$, the covariance increases for more negative $\beta$ value in central collisions but decrease in mid-central and peripheral collisions. We come back to this important observation later.

The panels in the right column of Fig.~\ref{fig:1} show the results of the $\rhon{2}$ and $\rhone{2}$. The $\beta$ dependence is more clearly revealed for these normalized quantities. In particular, we observe the strongest sensitivity in the UCC region, where a large positive $\beta$ leads to a negative $\rhon{2}$ and $\rhone{2}$, while a large negative $\beta$ increase them toward more positive direction. In the mid-central and peripheral regions, the values of $\rhon{2}$ and $\rhone{2}$ always decrease with increasing magnitude of $\beta$, independent of its sign.

The reason for the negative $\rhon{2}$ in central U+U collisions in the presence of large prolate deformation was clearly explained in Ref.~\cite{Giacalone:2019pca}. Denoting the radii for an ellipsoid as $R_a$, $R_b$ and $R_c$, then prolate deformation implies $R_a=R_b<R_c$. The ultra-central collisions correspond to events whose configurations are somewhere in between ``body-body'' collisions with long-axis parallel to each other in the transverse plane and ``tip-tip'' collisions with the long-axis parallel to the beam direction. Therefore, body-body collisions have large $\epsilon_2$ and large overlap area $A$, while the tip-tip collisions have small $\epsilon_2$ and small $A$. Such apparent anti-correlation between $\epsilon_2$ and $1/A$ naturally gives a strong anti-correlation between $v_2$ and $[\pT]$ and therefore negative $\rhon{2}$. The results in Fig.~\ref{fig:1} shows that the opposite is true in the presence of large oblate deformation for which $R_a=R_b>R_c$. In this case, body-body collisions have short-axis parallel to each other in the transverse plane, which are expected to give large $\epsilon_2$ and small overlap area $A$. Similarly, tip-tip collisions for oblate deformation have small $\epsilon_2$ but large $A$. Therefore, we expect an enhanced ``{\it positive}'' correlation between $\epsilon_2$ and $1/A$, leading to a stronger positive correlation between $v_2$ and $[\pT]$ observed in Fig.~\ref{fig:1}. This is quite interesting because the variance of $v_n$ and $[\pT]$ fluctuations do not distinguish between $\beta>0$ and $\beta<0$, while the $\rhon{2}$, being a three-particle correlator, can. Note that for maximum triaxiality deformation, for which $\gamma=\pi/6$ and $R_a<R_b<R_c$ forming an arithmetic sequence, there are no real distinction between ``body-body'' and ``tip-tip'' collisions, the deformation contribution is expected to reduce $\rhon{2}$ to zero. For general triaxiality $0<\gamma<\pi/3$, the signal is expected to interpolate between that for the oblate deformations and prolate deformation. We have checked this is indeed the case using a Glauber simulation of the initial-state estimator Eq.~\ref{eq:3}.

\begin{figure}[h!]
\begin{center}
\includegraphics[width=1.0\linewidth]{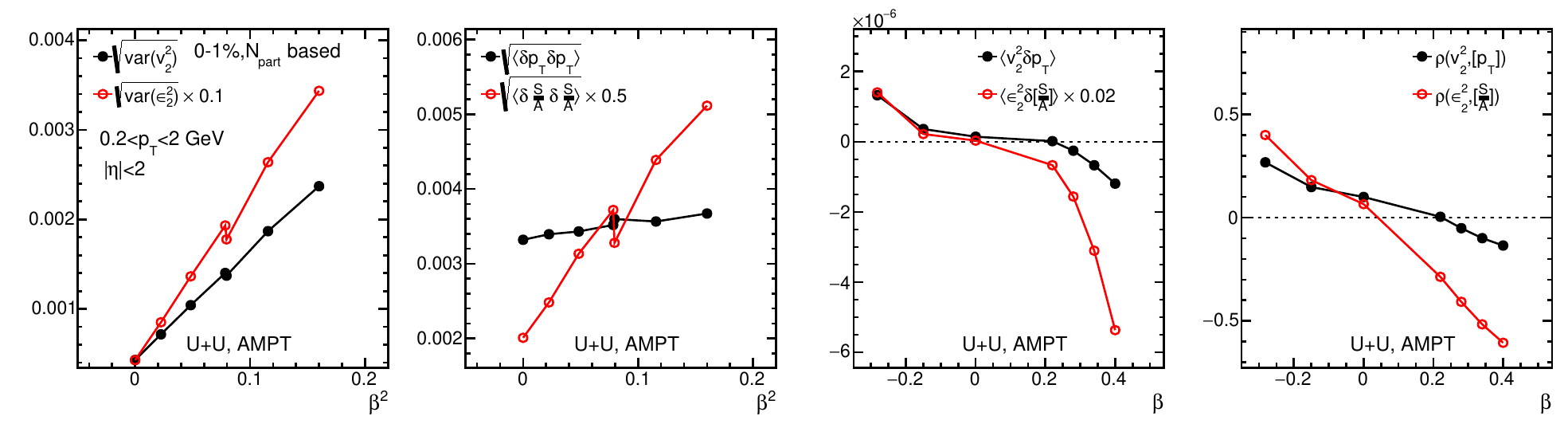}
\end{center}
\caption{\label{fig:3} The $\beta^2$ dependence of various quantities involved in $v_2-[\pT]$ correlation (solid circles) and $\varepsilon_2-S/A$ correlation (open circles) in top 0--1\% centrality based on the $\npart$ event class. The data points for $\beta=-0.28$ are shifted slightly to the right of the data points for $\beta=0.28$.}
\end{figure}

Next we would like to quantify the $\beta$ dependence observed in Figs.~\ref{fig:1}. We focus on the UCC region where the dependence on $\beta$ is strongest. We integrate the values for each observable in 0--1\% most central events, and plot them as a function either $\beta^2$ or $\beta$ in Fig.~\ref{fig:3}. Both $\sqrt{\var{2}}$ and $\sqrt{\varp}$ follow a nice linear increase with $\beta^2$. But the slopes of the increase are smaller than those for $\sqrt{\vare{2}}$ and $\sqrt{\vares}$, respectively. In contrast, the $\cov{2}$ and $\cove{2}$ show monotonic but non-linear decreases as a function of $\beta$. Interestingly, the normalized quantities $\rhon{2}$ and $\rhone{2}$ both approximately follow linear decrease as a function of $\beta$. Based on this, we obtain the following empirical approximation (more details see ~\cite{Jia:2021qyu}). 
\begin{align}\label{eq:8}
\cov{2}\approx a_1+(a_1+a_2\mathrm{Sign}(\beta))\beta^3\;,\; \rhon{2}\approx b_1+b_2\beta
\end{align}
$\mathrm{Sign}(\beta)$ ensure the increase of $\cov{2}$ with $|\beta|$ when $\beta<0$. For the range of $\beta$ value considered here, the cubic term is largely reduced by $\sqrt{\var{2}}$ and $\sqrt{\varp}$ an approximately linear dependence in $\beta$ for $\rhon{2}$. Very similar parameterization and observation can also be made about the initial-state quantities. The functional form for $\cov{2}$ and $\cove{2}$ can be thought as a event-by-event average of the product of a quadratic function $c_1+c_2\beta^2$ for $\sqrt{\var{2}}$ or $\sqrt{\vare{2}}$ and a linear function $d_1+d_2\beta$ for $[\pT]$ or $S/A$ (but with $\lr{c_2d_1}=0$ and $\lr{c_1d_2}=0$), and is generally expected for three-particle correlator if the signal for each particle has a linear dependence on $\beta$.

\begin{figure}[h!]
\begin{center}
\includegraphics[width=1.0\linewidth]{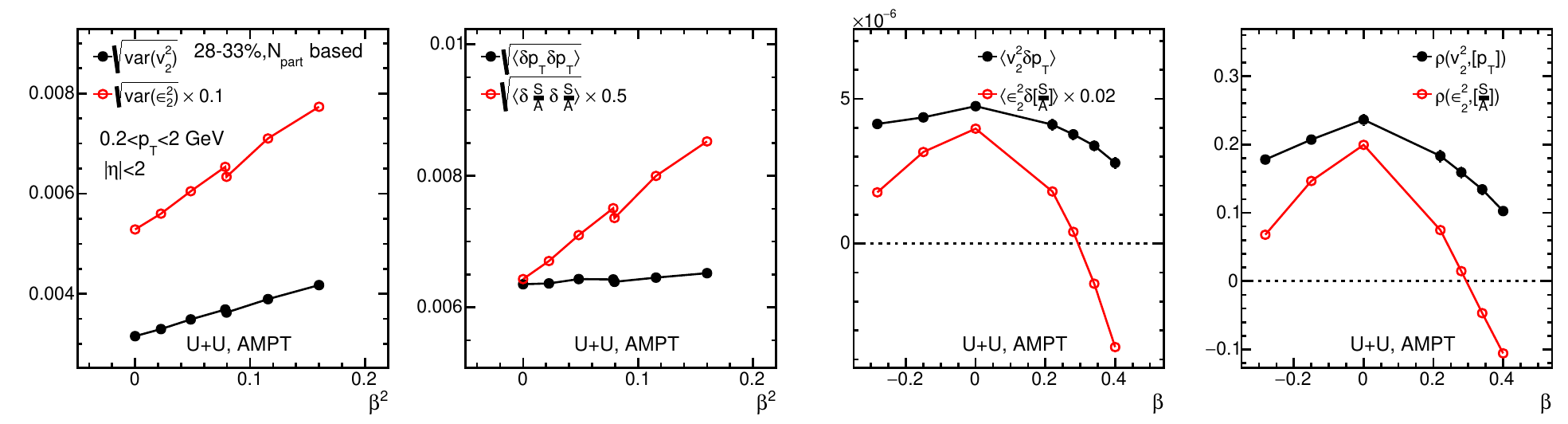}
\end{center}
\caption{\label{fig:4} The $\beta^2$ dependence of various quantities in $v_2-[\pT]$ correlation (solid circles) and $\varepsilon_2-S/A$ correlation (open circles) in 28-33\% centrality based on the $\npart$ event class. The data points for $\beta=-0.28$ are shifted slightly to the right of the data points for $\beta=0.28$.}
\end{figure}

Next, we perform the same analysis for mid-central 28--33\% collisions and results are shown in Fig.~\ref{fig:4}. Compared to central results, the $\cov{2}$ in the 28--33\% centrality shows a non-monotonic dependence on $\beta$, i.e it is largest for $\beta=0$ and decreases on both sides. Very similar qualitative dependence is also observed for the initial-state estimator $\cove{2}$. These trends are preserved for $\rhon{2}$ as shown in the right panel. We conclude that the $v_2-[\pT]$ correlation in mid-central collisions has rather complex dependence on $\beta$, which is driven entirely by $\epsilon_2-\frac{S}{A}$. Besides, this is the region where the centrality resolution effects plays an important role (see Fig.~\ref{fig:app2}), which need to be understood before we can draw strong physics conclusion by comparing with the experimental data. On the other hand, the results in 0--1\% have a more straightforward connection with collision geometry and they are insensitive to centrality resolution effects. Therefore, the UCC region is a sweet spot for experimental comparison to constrain the $\beta$ value and distinguish between prolate and oblate deformations. We expected this is true for hydrodynamic model in general. It would be interesting to also consider quadrupole deformation that exhibit triaxiality for which the three radii are different $R_a\neq R_b\neq R_c$. As discussed in Eq.~\ref{eq:1}, the general triaxiality is described by the triaxiality angle $\gamma$, which ranges from 0 for prolate shape to $\pi$/3 for oblate shape. It is natural to expect that the collisions of such system should have a $v_2-[\pT]$ signal in between the signal for prolate shape and oblate shape. Specifically, $\rhon{2}$ for highly deformed nuclei is expected to change sign from negative to positive, when $\gamma$ is varied from 0 to $\pi/3$. But the $v_2$ signal should be relatively insensitive to $\gamma$. A comparison of $v_2$ and $v_2-[\pT]$ for nuclei with similar mass number but different $\beta$ and triaxiality parameter values, say Lu and Hf region~\cite{Frauendorf:2013qca}, would be useful. 

Before closing the paper, we would like to discuss briefly two technical but important issues, which are presented in more details in Appendix~\ref{sec:app1}. Namely the influences of non-flow correlation and volume fluctuations. The influence of non-flow correlation were quantified by comparing results with those obtained from the subevents methods with $\eta$ gaps. We found that the non-flow correlations has no visible impact except in the very low $\nall$ region. To study the influence of volume fluctuations, the particles used to define event class (particles of centrality, POC) are chosen to be either similar or different from the particle used to calculate the $\rhon{2}$ (particles of interest, POI). When POC are chosen to have different $\pT$ or $\eta$ range from POI, the $\rhon{2}$ obtained for the same POI may differ significantly in mid-central and near-central collisions. However, the difference decreases for large $|\beta|$ value, indicating that the deformation induced contribution to $\rhon{2}$ is not affected by volume fluctuations. Most importantly, the $\rhon{2}$ in the UCC region, e.g. 0--1\%, is quite insensitive to the choice of POC. Therefore, the UCC region is clearly a sweet spot to isolate and constraint the influence of nuclear deformation to the $\rhon{2}$ (also the $v_2$ fluctuations as pointed out previously~\cite{Giacalone:2021udy}.).

\section{Summary}\label{summary}

We studied the influence of the nuclear quadrupole deformation on the fluctuations of harmonic flow $v_n$ and event-by-event mean transverse momentum $[\pT]$, and correlated fluctuations between $v_n$ and $[\pT]$ in Au+Au and U+U collisions at RHIC energy using the AMPT transport model. The variances of the $v_2^2$ and $[\pT]$ fluctuations,  $\sqrt{\var{2}}$ and $\sqrt{\varp}$, covariance $\cov{2}$ and Pearson correlation coefficient $\rhon{2}$ are calculated as a function of quadrupole deformation parameter $\beta$.  The $\sqrt{\var{2}}$ show a clear quadratic dependence on $\beta$ over a broad centrality range, driven by a similar quadratic $\beta$ dependence in the initial-state eccentricity $\varepsilon_2$. The $\sqrt{\varp}$ shows a very weak quadratic $\beta$ dependence, much weaker than the dependence observed for the variance of the size $R$ fluctuation in the initial state. This implies that AMPT model lacks a clear radial flow response to the overall system size fluctuations. 

The correlation between $v_2$ and $[\pT]$ shows a strong dependence on $\beta$. In the ultra-central collisions, $\rhon{2}$ shows a linear dependence $\beta$, i.e. it decreases for larger prolate deformation (more positive $\beta$) and increases for larger oblate deformation (more negative $\beta$). This is consistent with the expected correlation between $\varepsilon_2$ and radius of the system $R$ in the initial state for prolate and oblate nuclei in a simple geometrical picture. This finding shows great promise in using the strength and sign of $v_2^2-[\pT]$ correlation to constrain the triaxiality of the nucleus. In mid-central and peripheral region, the $\rhon{2}$ is largest for spherical nuclei and decrease for both prolate and oblate deformation. This behavior again follows qualitatively similar $\beta$ dependence for the expected correlation between $\varepsilon_2$ and $R$. The values of $\rhon{2}$ are not influenced by non-flow correlation, however they are sensitive to the choices of variable used to define event multiplicity in mid-central collisions due to centrality resolution and auto-correlation bias. Such dependence are found to be minimized in the ultra-central collisions where the results are found to be independent of event multiplicity definition. Therefore, $v_2-[\pT]$ correlation in UCC region can be used to provide direct connection to the initial collisions geometry and connect back to the shape of the colliding nuclei. Detailed comparison of the model prediction with the $v_2-[\pT]$ correlation data in Au+Au and U+U collisions should allow us to constrain the $\beta$ value of highly deformed U nucleus.

We thank Giuliano Giacalone for discussions and comments on the paper. This work is supported by DOE DEFG0287ER40331.
\clearpage
\appendix
\section{Influence of Non-flow correlations and volume fluctuations}\label{sec:app1}

The $v_n-[\pT]$ could have contributions that are unrelated to the initial-state geometry but arise from correlated particle production in the  momentum space such as jet fragmentation and resonance decays, known as ``non-flow''. The non-flow correlations can be suppressed by requiring pseudorapidity gaps between the particles in each triplet in the context of so-called standard, two-subevent and three-subevent methods~\cite{Jia:2017hbm}.  The influence of non-flow to $v_n-[\pT]$ correlation has been investigated in detail in Ref.~\cite{Zhang:2021phk} and was shown to be important only in the low multiplicity events. Here we repeat the same study also in the AMPT model. In the standard method used for the main results, all particles within $|\eta|<2$ are included.  In the two-subevent method, triplets are constructed by combining particles from two subevents labeled as $a$ and $c$ with a gap in between to reduce non-flow effects: $-2<\eta_a<-0.6\;,\; 0.6<\eta_c<2$. The two particles contributing to the flow vector are chosen as one particle each from $a$ and $c$, while the third particle providing the $\pT$ weight is taken from either $a$ or $c$. In the three-subevent method, three non-overlapping subevents $a$, $b$ and $c$ are chosen: $-2<\eta_a<-0.6\;,|\eta_b|<0.6\;,\; 0.6<\eta_c<2$. The particles contributing to flow are chosen from subevents $a$ and $c$ while the third particle is taken from subevent $b$. 

\begin{figure}[h!]
\begin{center}
\includegraphics[width=1.0\linewidth]{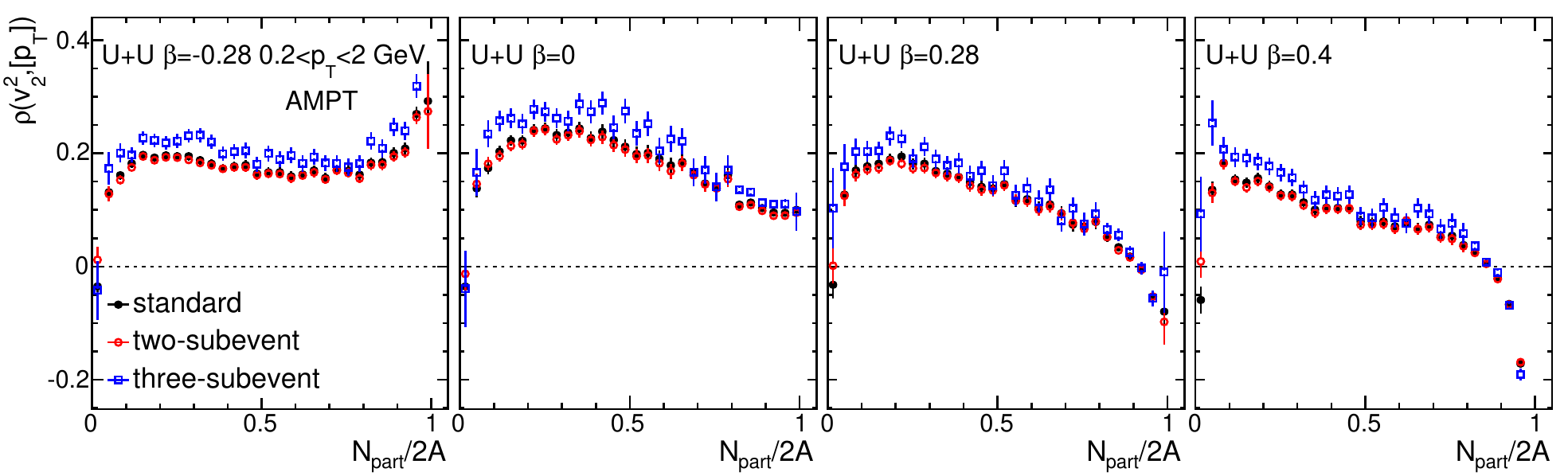}
\end{center}
\caption{\label{fig:app1} The $\npart$ dependence of $\rhon{2}$ obtained from the standard, two-subevent and three-subevent methods for U+U collisions for different $\beta$ values in each panel. From left to right they are $\beta=-0.28$, 0, 0.28 and 0.4. The event class used for averaging is based on $\npart$.}
\end{figure}

The comparison between these methods are shown in Fig.~\ref{fig:app1} for U+U collisions with several $\beta$ values. The two-subevent method agrees nearly perfectly with the standard method over broad $\npart$ range, while the results from the three-subevent method are systematically higher but the difference depends weakly on $\npart$. This difference suggests possible longitudinal decorrelation, which affects the strength of the correlation for particles separated in pseudorapidity~\cite{Jia:2014ysa,Jia:2017kdq}. The decorrelation effects were observed between the $v_n$ measured in two $\eta$ ranges~\cite{Khachatryan:2015oea,Aaboud:2017tql,Aad:2020gfz}, but in the present case the decorrelation between $v_n$ and $[\pT]$ might also play a role. We leave this topic to a dedicated study in the future.  

Our main analysis is performed using all hadrons with $0.2<\pT<2$ GeV and $|\eta|<2$, and the event centrality is defined using either $\npart$ or inclusive hadron multiplicity $\nall$ in $|\eta|<1$. In the calculation of various observables, the values obtained in each event are averaged over events with comparable multiplicity in $\nall$. They are then combined in broader multiplicity ranges of the event ensemble to obtain statistically more precise results. The event averaging procedure (also sometime referred to as centrality bin width correction) is necessary to reduce the effects of volume fluctuation within each event class definition~\cite{Zhou:2018fxx,Schenke:2020uqq,Bozek:2020drh}, but not completely eliminate it. This is because centrality for fixed value of $\nall$ is still smeared due to fluctuations in the particle production process. Since the $v_n$ and $[\pT]$ values vary with centrality, the smearing in centrality can lead to additional fluctuations of shape and size of the overlap region. Indeed, significant differences in $\rhon{n}$ were observed between centrality defined in mid-rapidity and centrality defined in the forward rapidity~\cite{ATLAS:2021kty}. To quantify the volume fluctuation effects, we performance a separate analysis. We calculate the $\rhon{n}$ using all hadrons with $0.2<\pT<2$ GeV and $|\eta|<1$ (Particles of Interest, POI), but using four different $\eta$ and $\pT$ choices for particles used to define event class (Particles of Centrality, POC): $\nall(1.5<|\eta|<2), \nall(|\eta|<1), \nall(1.5<|\eta|<2, \pT>0.5$ GeV), $\nall(|\eta|<1, \pT>0.5$ GeV). Note that particles used to define $\nall (1.5<|\eta|<2)$ have no overlap with particle used to calculate $\rhon{n}$, which could reduce the auto-correlation effects associated with non-flow.

\begin{figure}[h!]
\begin{center}
\includegraphics[width=1.0\linewidth]{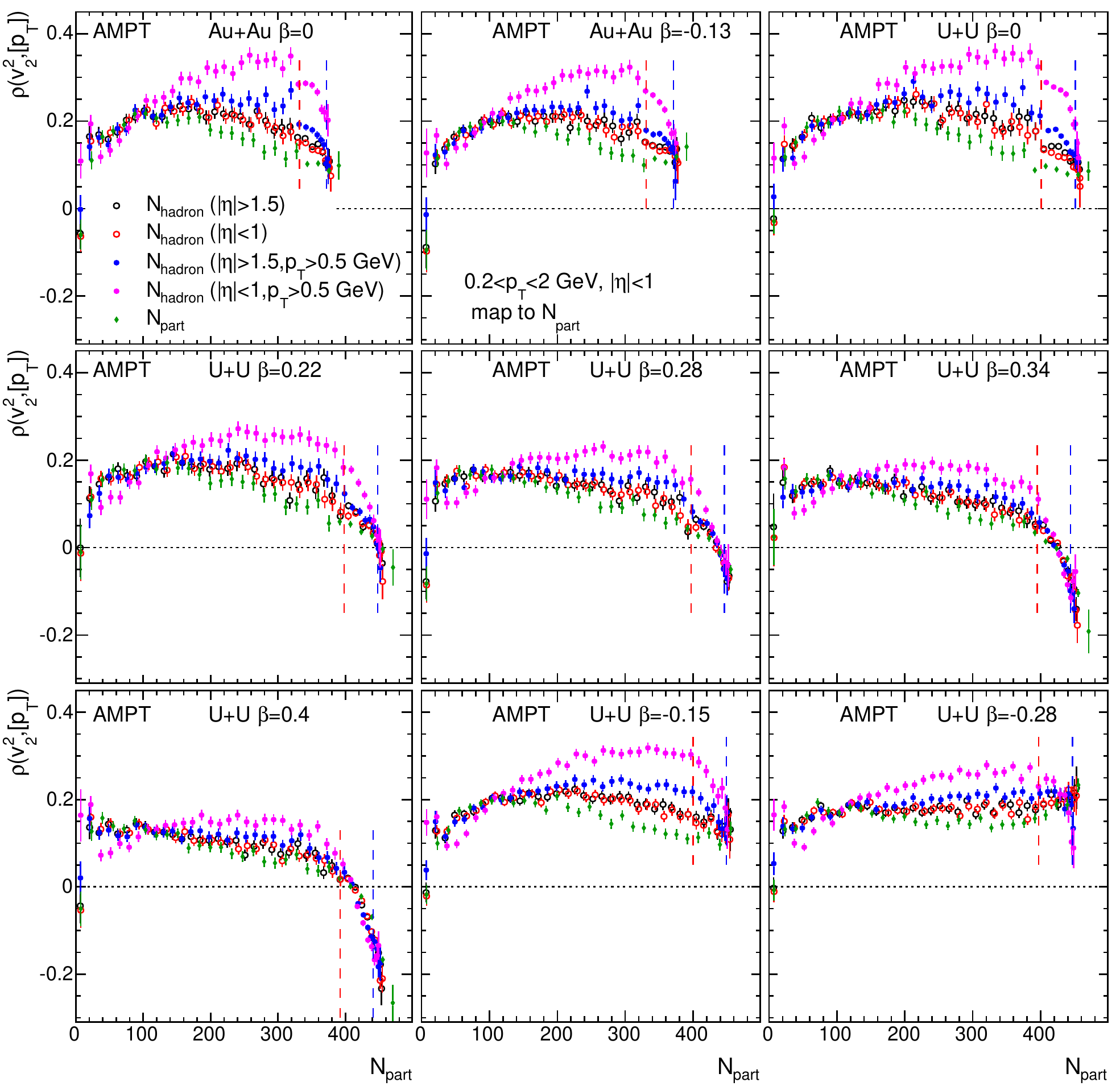}
\end{center}
\caption{\label{fig:app2} The $\npart$ dependence of $\rhon{2}$ calculated for hadrons in $0.2<\pT<2$ GeV and $|\eta|<1$ but compared between five different event class definitions after mapping to the common $\npart$ $x$-axis. They are shown separately for Au+Au or U+U collisions with various $\beta$ values as indicated in each panel. The vertical dashed lines indicate the locations for the 1\% and 5\% highest $\npart$ values.}
\end{figure}
In Fig.~\ref{fig:app2}, we demonstrates the sensitivity to volume fluctuation by comparing the results of $\rhon{2}$ for four different POC. They are mapped to $\npart$ and compared with the results obtained directly by binning events according to $\npart$.  Each panel shows the results obtained for Au+Au or U+U collisions with a particular $\beta$ value. The results for event class based on $\npart$ always has the smallest $\rhon{2}$ values, except in the very peripheral region. Results obtained for different $\nall$ event classes show large spread in mid-central collisions, but they converge to a common results in the most central region and cross each other at $\npart\sim80-120$ depending on $\beta$, which corresponds approximately to the average $\npart$ for minimum-bias Au+Au or U+U collisions.  The spreads between different $\nall$ are largest for $\beta=0$ and reduce significantly for large $|\beta|$ values, implying that the deformation contribution is not affected by centrality resolution. Looking in each panel in more detail, we find that the results based on $\nall$ without $\pT$ cut are very close to each other, albeit still a bit higher than $\npart$ case. On the other hand, the results based on $\nall$ that include only high $\pT$ hadrons, i.e. $\pT>0.5$ GeV are much larger. 

 The largest enhancement is observed for $\nall(|\eta|<1, \pT>0.5~\mathrm{GeV})$, but only a slight increase is observed for $\nall(|\eta|>1.5, \pT>0.5~\mathrm{GeV})$. We conclude that the when only high $\pT$ hadrons are counted for $\nall$, there is a large auto-correlation between the $\nall$ and $\rhon{2}$ if the $\eta$ ranges for these two quantities have overlap. Such auto-correlation is minimal, however, if $\nall$ definition also include low $\pT$ particles. The results from STAR Collaboration are based on number of charged particles with $|\eta|<0.5$ and $\pT>0.2$ GeV for event class and $|\eta|<1$ and $\pT>0.2$ GeV for calculation of $\rhon{2}$~\cite{jjia}. In this case, we expect the auto-correlation effects are rather modest. The results from ATLAS Collaboration were based on $|\eta|<0.5$ and $\pT>0.5$ GeV for both $\rhon{2}$ and event class definition. The auto-correlation bias could be less severe since the $\lr{\pT}$ is larger at LHC and $\pT>0.5$ GeV selection should already included most of the hadrons. This point certainly deserves a dedicated study. But independent of the finding, we would like to emphasize that the results in the most central collisions, e.g. 0--1\% centrality as indicated by the vertical dashed line each panel, are rather stable against auto-correlation and/or the volume fluctuation effects. In the presence of large deformation, the stable region increases further and may extend to top 0--5\% centrality. Therefore, the sign-change region of $\rhon{2}$ in the U+U collisions observed by the STAR Collaboration~\cite{jjia}, covering the top 0--8\% most central events, should be rather robust and independent of the particular choice centrality used. 
\clearpage
\section{Results calculated with $\nall$ event class and plotted as a function of $\nall$}\label{sec:app2}
\begin{figure}[h!]
\begin{center}
\includegraphics[width=1.0\linewidth]{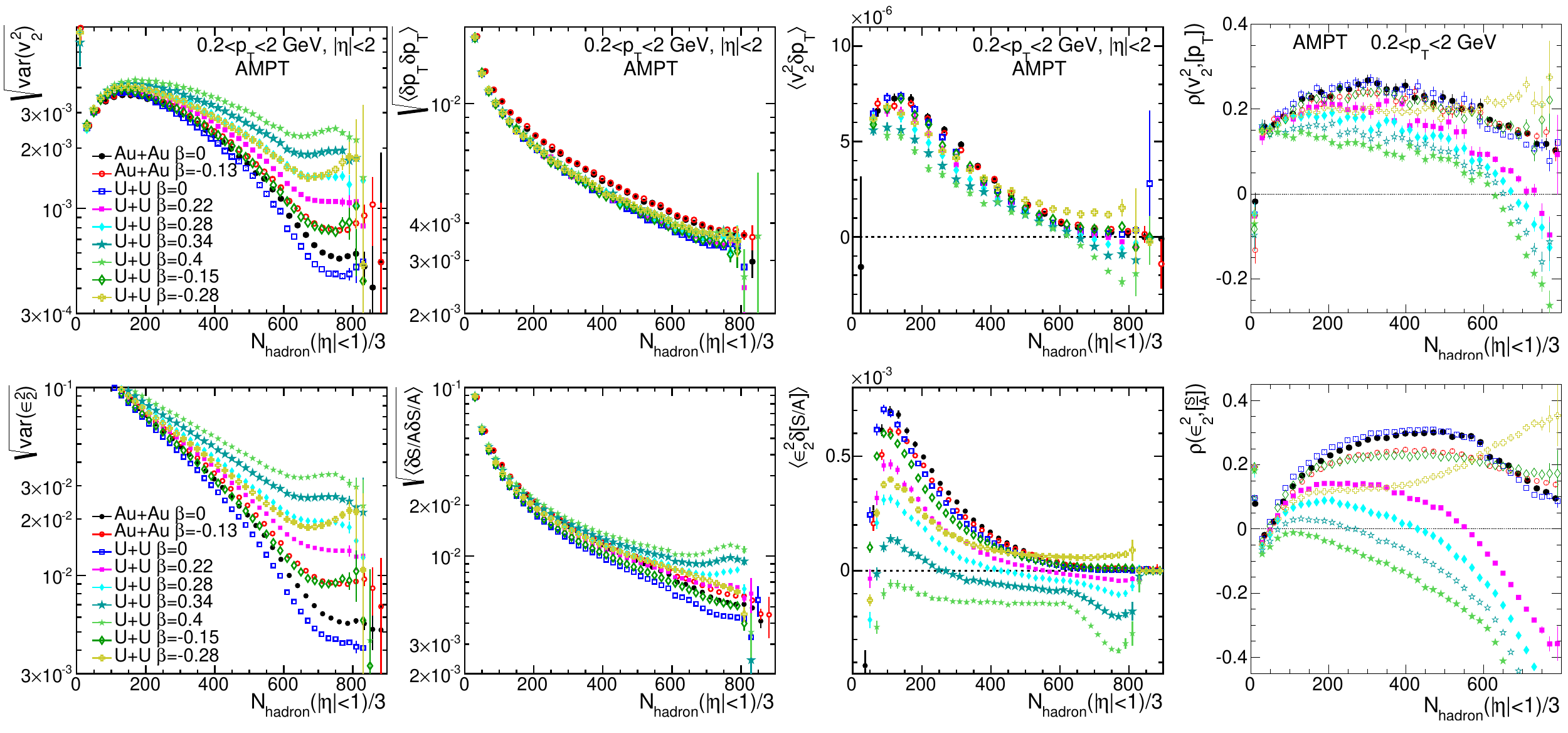}
\end{center}
\caption{\label{fig:app3} The $\nall(|\eta|<1)$ dependence of $v_2^2$ variance (top-left), $[\pT]$ variance (2nd from top-left), flow-$[\pT]$ covariance $\cov{2}$ (3rd from top-left) and Pearson coefficients $\rhon{2}$ (top-right)  in Au+Au and U+U collisions with different deformation parameter $\beta$. The bottom panels show the corresponding initial-state estimators in the Glauber model. The event class used for averaging is based on $\nall(|\eta|<1)$.}
\end{figure}

So far, the results are either calculated with $\npart$ or $\nall$ as POC, but are always presented as a function of $\npart$. Note that, in the second case, the $\npart$ is defined as the average $\npart$ for events with given $\nall$. Due to relative smearing, events in top 0--1\% of $\nall$ are different from events in top 0--1\% of $\npart$, in particular the $\lr{\npart}$ for events in top 0--1\% of $\nall$ is generally smaller than the $\lr{\npart}$ for events in top 0--1\% of $\npart$. This simply means that the influence of volume fluctuations for events in top 0--1\% of $\nall$ is generally larger for events in top 0--1\% of $\npart$ (see Fig.~\ref{fig:app2}).  

\begin{figure}[h!]
\begin{center}
\includegraphics[width=1.0\linewidth]{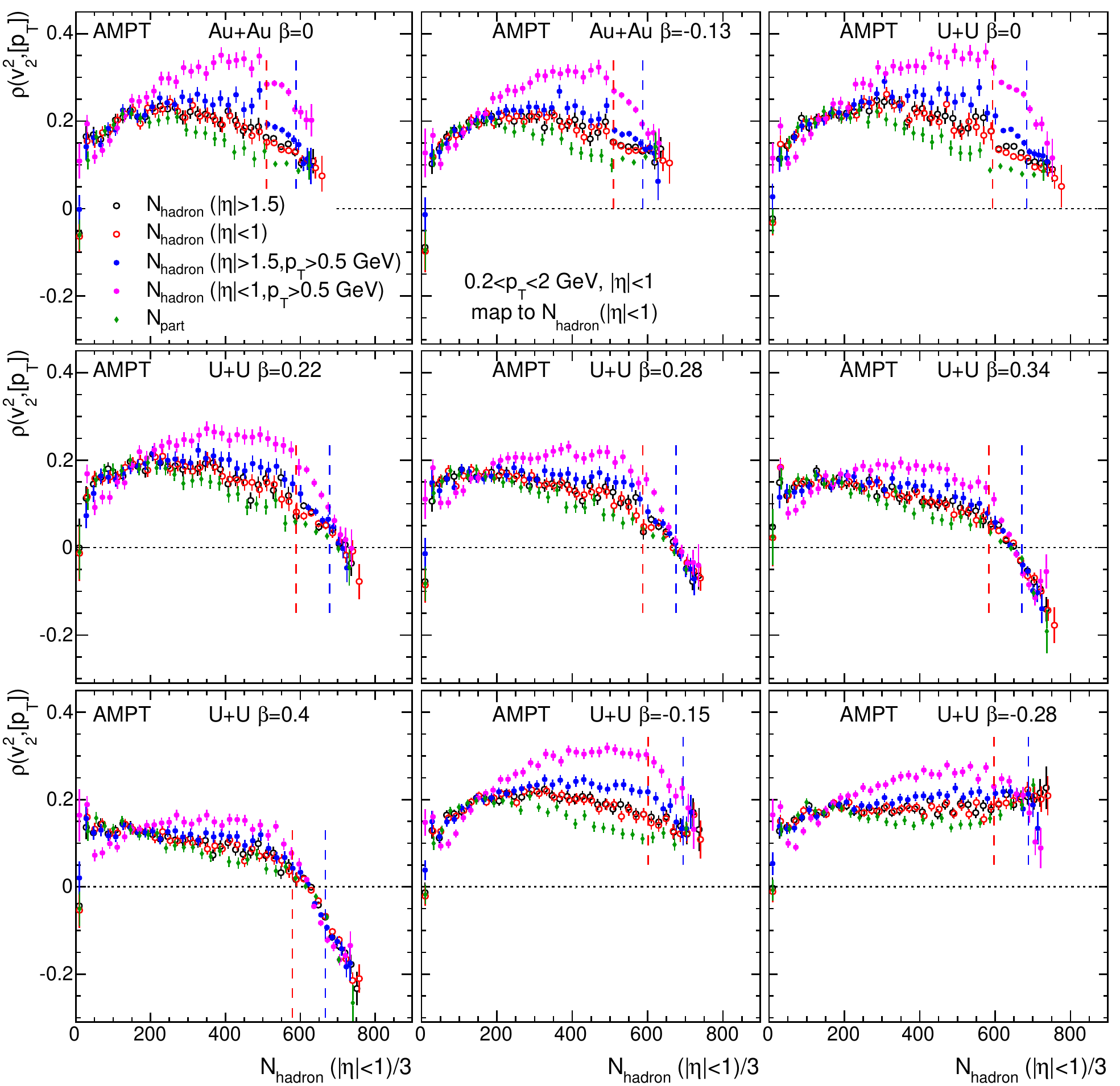}
\end{center}
\caption{\label{fig:app4}  The centrality dependence of $\rhon{2}$ calculated for hadrons in $0.2<\pT<2$ GeV and $|\eta|<1$ and compared between give different event class definitions after mapping to the common $x$-axis defined by $\nall(|\eta|<1)/3$.  They are shown separately for Au+Au or U+U collisions with various $\beta$ values as indicated in each panel. The vertical dashed lines indicate the locations for the top 1\% and 5\% highest $\nall$ values.}
\end{figure}

Figure~\ref{fig:app3} shows the components of Pearson correlation coefficients calculated with final-state hadrons (top row) and the initial-state participating nucleons (bottom row). They are calculated using $\nall$ as POC for event averaging and plotted as a function of $\nall$. They should be contrasted to Fig.~\ref{fig:1}.  The main features are similar, except that the distributions are much more smeared in the UCC region. Figure~\ref{fig:app4} shows the influence of volume fluctuation with the POC defined in the same way as in Fig.~\ref{fig:app2}. In fact, they are exactly the same data points, and the only difference is they are mapped to $\nall$. One noticeable difference from Fig.~\ref{fig:app2}, however, is that the influence of volume fluctuations is still visible in the top 0--1\% centrality based on $\nall$. One need to use even more extreme selection, such as top 0--0.2\% range for $\nall$ in order to minimize the volume fluctuation effects. 

\begin{figure}[h!]
\begin{center}
\includegraphics[width=1.0\linewidth]{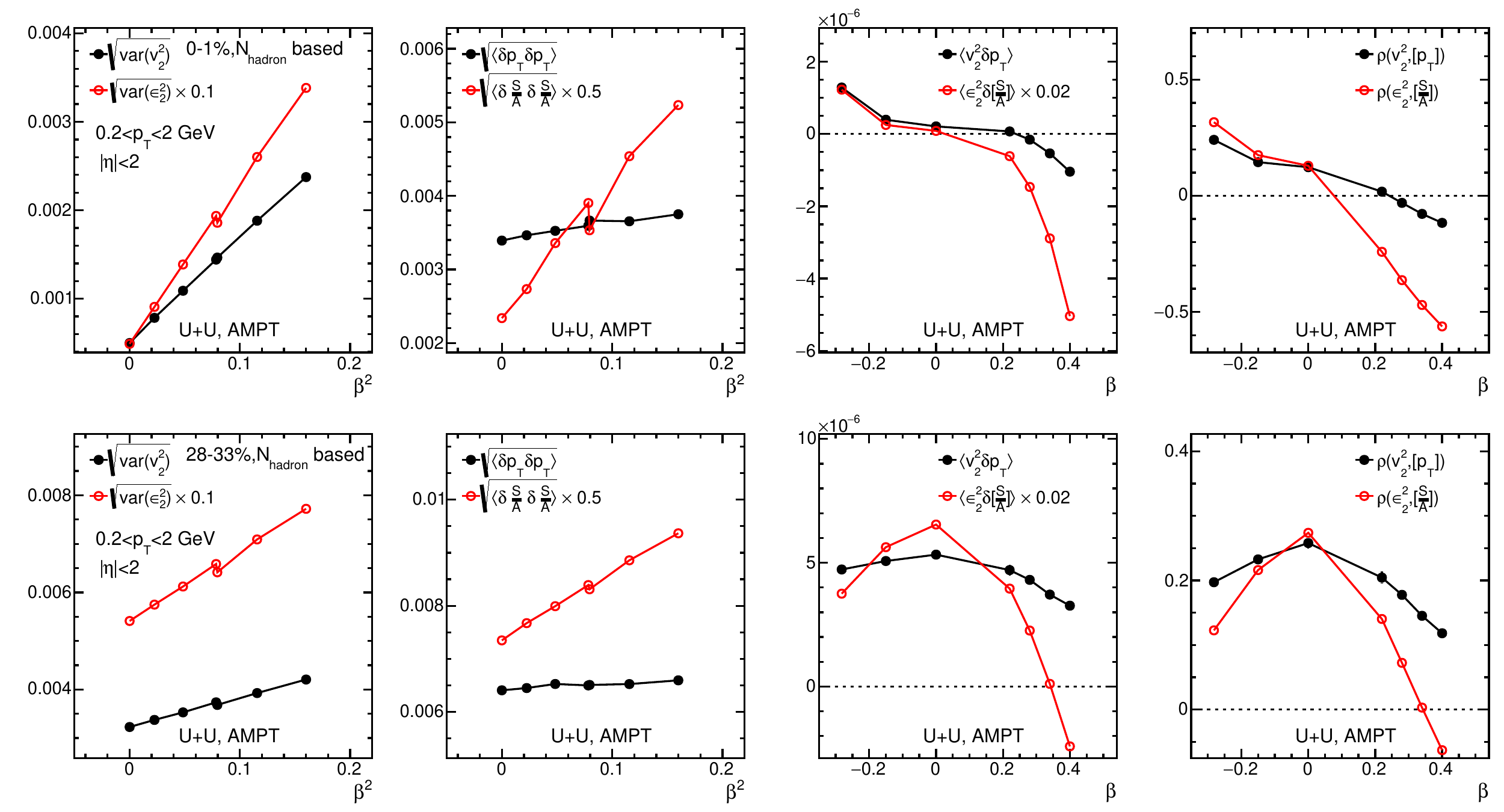}
\end{center}\vspace*{-0.7cm}
\caption{\label{fig:app5} The $\beta^2$ dependence of various quantities for the $v_2-[\pT]$ correlation (solid circles) and the $\varepsilon_2-S/A$ correlation (open circles) in the top 0--1\% centrality (top row) and 28--33\% centrality (bottom row) based on the $\nall(|\eta|<1)$ event classes. The data points for $\beta=-0.28$ are shifted slightly to the right of the data points for $\beta=0.28$.}
\end{figure}

Finally, Fig.~\ref{fig:app5} shows the $\beta$ dependence of the components for $v_2-[\pT]$ in the 0--1\% (top row) and the 28--33\% (bottom row) centrality based on $\nall$. There are some quantitively differences when comparing to Figs.~\ref{fig:3} and \ref{fig:4}. but the qualitative behaviors are similar. The differences are particularly noticeable for initial-state quantities since these quantities are more directly connected to $\npart$ than to $\nall$. But the differences for the final-state quantities are much smaller. 

\section{Results for higher order harmonics}\label{sec:app3}
\begin{figure}[h!]
\begin{center}
\includegraphics[width=0.6\linewidth]{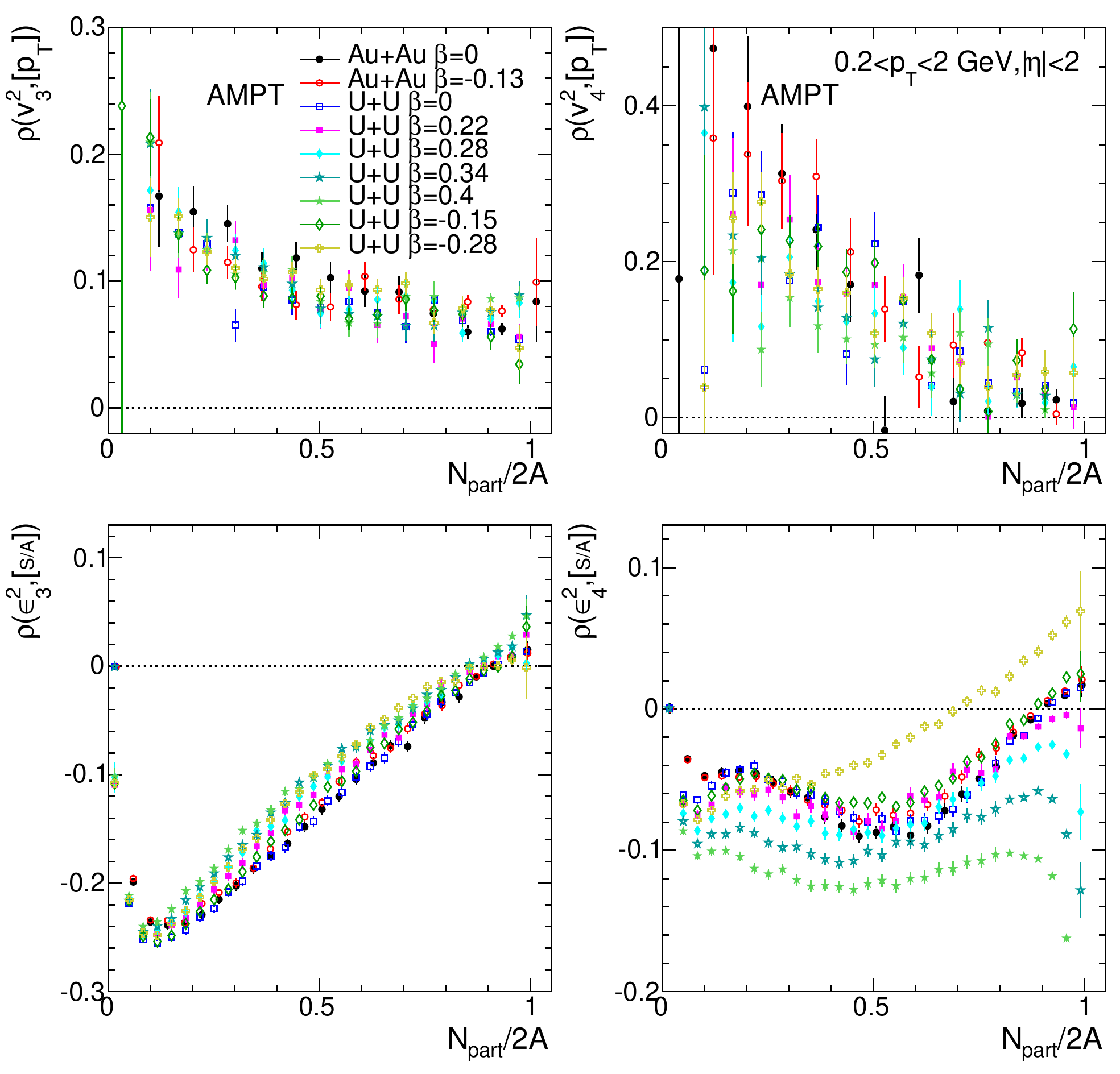}
\end{center}\vspace*{-0.7cm}
\caption{\label{fig:app6} The $\npart$ dependence of $\rhon{3}$ (top-left) and $\rhon{4}$ (top-right) in Au+Au and U+U collisions with different deformation parameter $\beta$. The bottom panels show the corresponding initial-state estimators in the Glauber model. The event class used for averaging is based on the $\npart$.}
\end{figure}

Figure~\ref{fig:app6}  show the results of Pearson correlation coefficients for higher-order flow harmonics $n=3$ and $n=4$. Within the present statistics precision, we conclude that $\rhon{3}$ and $\rhon{4}$ are relatively insensitive to the quadrupole deformation effects. We observe, however, that the corresponding initial-state quantities $\rhone{3}$ and $\rhone{4}$ show qualitatively very different $\npart$ dependence shape. This behavior implies that either they are not good initial-state estimators for $v_3-[\pT]$ and $v_4-[\pT]$ correlations, or correlation between higher-order flow and $[\pT]$ are dominated by the dynamical effects in the final state. An alternative estimator based on ratio of total energy and total entropy was found to reproduce qualitatively the $\rhon{3}$ data from ATLAS~\cite{Giacalone:2020awm}.

\vspace*{-0.7cm}

\clearpage
\bibliography{amptvnpt}{}
\bibliographystyle{apsrev4-1}

\end{document}